\documentclass[a4paper]{article}
\usepackage[T1]{fontenc}
\usepackage{graphicx}
\usepackage{amsmath}
\usepackage{hyperref}
\usepackage{todonotes}         
\usepackage{tikz}              
\usetikzlibrary{shapes,arrows} 
\usepackage[english]{babel}    
\usepackage{enumitem}          
\usepackage{cite}              
\usepackage{xspace}            
\usepackage{semantic}
\usepackage{array}             
\usepackage[edges]{forest}     
\usepackage{multicol}
\usepackage{amsfonts}
\usepackage{stmaryrd}

\newcommand{\pun}{\textsf{pun}\xspace}
\newcommand{\fun}{\textsf{FUN}\xspace}
\newcommand{\quickcheck}{QuickCheck\xspace}

\usepackage{listings}

\newcommand{\haskell}[1]{\lstinline{#1}}

\begin{document}

\newcommand{\forceindent}{\leavevmode{\parindent=1em\indent}}

\def\Name{\ensuremath{\text{\bf Name}}}
\def\OR{\ensuremath{\ |\ }}
\def\TO{\ensuremath{\rightarrow}}
\def\FROM{\ensuremath{\leftarrow}}
\def\LB{\ensuremath{\llbracket}}
\def\RB{\ensuremath{\rrbracket}}

\newcommand \BX [1]
  {\scriptsize\framebox{{\raisebox{0pt}[0.7\baselineskip][0.01\baselineskip]{\small #1}}}}

\newcommand\Axiom[2]
                 {\ensuremath{\text{\small #1}:\frac{\displaystyle}
                 {\displaystyle #2}}
                 }
\newcommand\InfOne[3]
                 {\ensuremath{\text{\small #2}:\frac{\displaystyle #1}
                 {\displaystyle #3}}
                 }
\newcommand\InfTwo[4]
                 {\ensuremath{\text{\small #3}:\frac{\displaystyle #1 \quad #2}
                 {\displaystyle #4}}
                 }
\newcommand\InfThree[5]
                 {\ensuremath{\text{\small #4}:
                     \frac{\displaystyle #1 \quad #2 \quad #3}
                          {\displaystyle #5}}
                 }
\newcommand\InfFour[6]
                 {\ensuremath{\text{\small #5}:
                     \frac{\displaystyle #1 \quad #2 \quad #3 \quad #4}
                          {\displaystyle #6}}
                 }
\newcommand\InfFive[7]
                 {\ensuremath{\text{\small #6}:
                     \frac{\displaystyle #1 \quad #2 \quad #3 \quad #4 \quad #5}
                          {\displaystyle #7}}
                 }

\newcommand\INT[0]{\textbf{int}}
\newcommand\BOOL[0]{\textbf{bool}}
\newcommand\REC[0]{\textbf{rec}\ }
\newcommand\FST[1]{\textbf{fst}(#1)}
\newcommand\SND[1]{\textbf{snd}(#1)}
\newcommand\LAMBDA[2]{\lambda #1 . #2}
\newcommand\LET[3]{\textbf{let}\ #1 = #2\ \textbf{in}\ #3}
\newcommand\IF[3]{\textbf{if}\ #1\ \textbf{then}\ #2\ \textbf{else}\ #3}
\newcommand\CASE[4]{\textbf{case}\ #1\ \textbf{of}\ ; \textbf{leaf} \to #2\ ; #3 \to #4}
\newcommand\NODE[4]{[\ \textbf{node}\ #1\ #2\ #3\ #4\ ]}
\newcommand\GEN[2]{{\LB #1 \RB}_{#2}}

\title{\pun: Fun with Properties;\\Towards a Programming Language With
  Built-in Facilities for Program Validation}

\author{
Triera Gashi$^1$ \and
Sophie Adeline Solheim Bosio$^1$ \and
Joachim Tilsted Kristensen$^1$ \and
Michael Kirkedal Thomsen$^{1,2}$}

\date{
  $^1$ University of Oslo, Oslo, Norway\\
  $^2$ University of Copenhagen, Copenhagen, Denmark
}
\maketitle              
\begin{abstract}
Property-based testing is a powerful method to validate program correctness. It is, however, not widely use in industry as the barrier of entry can be very high.
One of the hindrances is to write the generators that are needed to generate randomised input data. Program properties often take complicated data structures as inputs and, it requires a significant amount of effort to write generators for such structures in a invariant preserving way.

In this paper, we suggest and formalise a new programming language \pun; a simple functional programming with properties as a built-in mechanism for program validation. We show how to generate input for \pun properties automatically, thus, providing the programmer with a low barrier of entry for using property-based testing. We evaluate our work a on library for binary search trees and compare the test results to a similar library in Haskell.
\end{abstract}

\section{Introduction}
To reduce the risk of
defects in modern software, it is common to perform a form of validation that ensure the software is built according to its specification~\cite{Graham:2022}.
As an example, software developers may rely on testing for providing
evidence that their software behaves as intended, and thereby increase the
likelihood that their work is correct.  However, software testing can be
hard and cumbersome and, thus, is not a very popular activity among
developers~\cite{Whittaker:2000,Hughes:2016}.

Property-based testing is a testing methodology that supports comprehensive
software testing. When compared to unit and integration testing,
property-based testing may justify greater confidence in program
correctness; Instead of a handful of inputs and their expected output, the
programmer provides a set of properties/assertions, that should hold for a
specific program fragment. Idealised, a tool then automatically generates
random test inputs and checks that these properties hold. This way, the
programmer is relieved of coming up with inputs and predicting the
outcome. Each newly generated test input increases confidence, rather than
the same hundred test cases being run over again. As the so-called
\emph{pesticide paradox} states, running the same test cases repeatedly will
not find new bugs~\cite{Graham:2022}.

A popular tool for performing property-based testing is called
\quickcheck~\cite{ClaessenHughes:2000:ACM}. \quickcheck is a potent
combinator library capable of generating test cases based on
\emph{assertions} and \emph{test input generators}.
Although \quickcheck is a well-known testing tool (especially in the
academic environment), a significant hindrance to its adoption in industry
settings is the need for handwritten
generators~\cite{LampropoulosEtal:2017}.  In practice the programmer not
only needs to write the properties, but also a set of generators that can
generate random inputs for the tests.  Generating inputs for tests
\quickcheck can do out-of-the-box, but for more complex user-defined data
types and properties, the programmer is required to write a generator for
the data type by hand.

In this work we want to alleviate the burden of writing generators.
For this, we propose \pun, a functional programming language with a built-in construct for defining and checking properties.
This can in turn be used to automatically generate test input generators, to facilitate the use of property-based testing.
Automatic program generation is a hard problem (and in general undecidable), so it is expected to come at the cost of the quality compared to good generators.
However, with the conjecture that any testing is better than none, this can still give a significant improvement.

To give a better feeling of the problem of implementing generators, consider
for following example. We will look at the built-in addition operator,
\haskell|+|, where a property we may want to test, is commutativity. Here
the \pun program might be as simple a single line:

\begin{lstlisting}[escapeinside=\%\%]
property add-is-commutative m n . m + n == n + m.
\end{lstlisting}
\noindent
The property states that any two \pun terms \haskell|m| and \haskell|n|,
will satisfy the above equation.  The addition operation in \pun is only
defined for terms of type integer. So, in order to check it, we must be able to
generate two arbitrary terms of type \haskell|integer| and substitute these
terms into the equation. Evaluating the resulting term will have type
\haskell|boolean|, and the property holds if the term evaluates to
\haskell|true| for any choice of \haskell|m| and \haskell|n|.

In this case, since \quickcheck comes with a reasonable integer generator
out of the box, it is relatively easy to generate input data for the
property -- but for more complex properties and types, this is not the
case. Take for instance the \pun property
\begin{lstlisting}[escapeinside=\%\%]
property plus-zero-identity f x . f (x + 0) == (f (x)) + 0 .
\end{lstlisting}
\noindent
where there are several choices to be made. How would you write generators
for the terms \haskell|f| and \haskell|n|? -- In both of the above examples,
\pun automatically generates the appropriate closed terms and substitutes
them into the term in question. And \pun trusts that the property holds if
it holds for several (currently 50) choice of subterms (\haskell|m| and
\haskell|n| or \haskell|f| and \haskell|x|), and it communicates this fact
by outputting dot per test passed.
\begin{lstlisting}[escapeinside=\%\%]
testing plus-commutes: ............................................... ok
testing plus-zero-identity: .......................................... ok
\end{lstlisting}

Suppose now, that we also want to check if subtraction is commutativity and write the program
\begin{lstlisting}[escapeinside=\%\%]
property sub-is-commutative m n . m - n == n - m.
\end{lstlisting}
\noindent
Again, our property checker will generate the appropriate terms and do the
substitution, The only difference in this case, is that \pun will output the
test case for which the property did not hold.
In this example, that would be the first term
where the generated terms are non-equal integer terms, for instance
\begin{lstlisting}[escapeinside=\%\%]
testing subtraction-commutes: .."failed with counter example :"
  ((\ x -> x + 2) 3) - 7 == 7 - ((\ x -> x + 2) 3)
"after 2 tests"
\end{lstlisting}

In order to save the programmer the work of writing generators by hand, we
need to write two functions.  One function that, given information about the
program bindings, can generate a generator of the appropriate type. A second
function that uses those generators and substitutes the generated terms
into the property. The problem of generating these two functions will be the
focus of this paper.

A re-implementation of \quickcheck is Luck~\cite{LampropoulosEtal:2017}, a domain-specific language that in principle has
the same goal as our work.
However, the way that writing generators is made easier in Luck is by
decorating predicates with lightweight annotations, while we take a more
general approach.

\paragraph{Structure:}
Section~\ref{sec:approach} will give a larger example that details our approach. Section~\ref{sec:pun-lang} formalises the syntax and type system of \pun. Section~\ref{sec:gengen} describes how to generate generators from \pun programs. Section~\ref{sec:interesting-terms} discusses how to limit the generators for cases where inputs quickly diverges, while Section~\ref{sec:evaluation} evaluated our approach. Finally, in Section~\ref{sec:conclusion} we conclude the work.

The Haskell implementation of \pun, and the benchmarks can be found at \url{https://github.com/jtkristensen/pun-lang}.

\section{Approach to the Work}
\label{sec:approach}
While illustrative, the examples above are too small sufficiently
demonstrate the problems with generators. Therefore, we will in
the following detail \pun on a larger example.
In~\cite{Hughes:2019},
Hughes describes different techniques for writing good properties for pure functions in Haskell. As an example, he uses binary search trees (BSTs) and five common
operations on these, and uses QuickCheck to test different properties for each
operation.
In the Haskell code he needs a hand-written generator to test on the BST.

Below, we have implemented the example in \pun and Haskell for comparison.

\begin{lstlisting}[escapeinside=\%\%]
insert : integer -> integer
       -> (bst integer integer -> bst integer integer) .
insert k1 v1 t =
  case t of
  ; leaf      -> [node leaf k1 v1 leaf]
  ; [node l k2 v2 r] ->
    if equal k1 k2
    then [node l k2 v1 r]
    else if   k1 <= k2
         then [node (insert k1 v1 l) k2 v2 r]
         else if   k1 > k2
              then [node l k2 v2 (insert k1 v1 r)]
              else [node (leaf) k1 v1 (leaf)] .

property insert-valid k v t .
  if   valid t
  then valid (insert k t)
  else true .

property find-post-present k v t .
  find_equal (find k (insert k v t)) ([node leaf k v leaf]) .
\end{lstlisting}

\noindent
Given that we have a function that checks whether a tree is a valid binary
search tree, we can formulate a property that inserting a key
value pair should result in a valid BST. This is what Hughes
describes as a \emph{validity} property. Another property to formulate is that you
should be able to find a key after inserting it, which is a post-condition property.
The way we have implemented the \haskell|find| function in \pun, is to return a leaf if it
cannot find the key, otherwise it returns the key stored in a node.

The equivalent Haskell implementation looks like so:

\begin{lstlisting}[escapeinside=\%\%]
insert :: Ord k => k -> v -> BST k v -> BST k v
insert k v Leaf = Branch (Leaf) k v (Leaf)
insert k v (Branch left k' v' right)
    | k == k' = Branch left              k' v  right
    | k <  k' = Branch (insert k v left) k' v' right
    | k >  k' = Branch left              k' v' (insert k v right)
insert k v _  = Branch (Leaf) k v (Leaf)

instance (Ord k, Arbitrary k, Arbitrary v) => Arbitrary (BST k v) where
arbitrary = do
  kvs <- arbitrary
  return $ foldr (uncurry insert) nil (kvs :: [(k, v)])
shrink = filter valid . genericShrink

prop_InsertValid :: Key -> Val -> Tree -> Bool
prop_InsertValid k v t = valid (insert k v t)

prop_FindPostPresent :: Key -> Val -> Tree -> Property
prop_FindPostPresent k v t = find k (insert k v t) === Just v
\end{lstlisting}

When comparing the \pun and Haskell implementation, one can see that the
properties themselves are similar (aside from differences in syntactic
sugar). The big difference is providing the \haskell|Arbitrary| instance,
which tells \quickcheck how to generate an arbitrary BST. One is also
presented with the option of writing a shrink a tree in order to give the
smallest possible failing BST that makes a property fail. In \pun, the user
does not need to provide either, thus trading off the potential quality of
writing a good generator by hand for the ergonomics of getting one for free.

There are several things to consider when writing a generator, to ensure that
the generator generates intended terms. Hughes uses
a \haskell|valid| function in his arbitrary instance to make sure that only valid
BSTs are generated. When writing generators you will often have to write similar functions
for the arbitrary instance to ensure that what you generate has the desired
properties.
These functions must
then also be correct and tested themselves before being used in the arbitrary instance.
It is also difficult to make sure that the generator generates
terms of useful sizes. Often you have to use combinators such as \haskell|frequency| to increase
the probability of something being generated. You need to investigate what your
generator generates, and see if it produces the desired results.
In all, there is much work required to write a generator, and none of these
tasks are easy.

There is, however, an issue with not having the option of modifying, putting
constraints or defining predicates about what is being generated. In Hughes'
arbitrary instance, he uses \haskell|insert| in order to generate trees that
are ordered correctly, with smaller keys in the left subtree and greater keys in
the right subtree.
He also specifies that shrinking a BST should result in a valid
BST.
Our generator will generate any pairs of keys and values to the tree, so
there is a high probability that the tree generated does not inhabit this
BST property. There are two options in this case; one could
define weaker properties, such as the one above that only checks that the
insert property holds for the generated trees that are valid, but then the
question is how often do we randomly generate valid trees?
Another option is to transform the output of the generator so that it has
this property, by for instance writing a function \haskell|validify| that
takes the randomly generated tree and gives the correct structure:

\begin{lstlisting}[escapeinside=\%\%]
property insert-valid k v t . valid (insert k v (validify t)) .

validify : bst integer integer -> bst integer integer .
validify t =
  case t of
  ; leaf           -> leaf
  ; [node l k v r] -> insert k v (union (validify l) (validify r)) .
\end{lstlisting}

It is difficult to generate algebraic data types that must have a certain structure,
unless you are able to modify the generated terms to have the desired properties.
Even though the programmer does not need to provide a generator, they might have
to provide a function such as \haskell|validify|, and it is worth noting that
this can be as difficult to write as the generator itself.

\section{The \pun Explained in Detail}
\label{sec:pun-lang}
The language \pun is a higher-order functional language
that uses call-by-value parameter passing. \pun is furthermore extended with
built-in facility for property-based testing.

\pun is a garden variety functional programming language, inspired by simple
functional programming languages such as \fun\footnote{Our version of \pun
  is actually closer related to \fun from Andrzej Filinski's unpublished
  lecture notes on formal semantics and types.} from
Pierce~\cite{Pierce:2002:book}, and REC from Winskel~\cite{Winskel:1993}.
As such, this section omits the details with the operational semantics in
the interest of saving space for discussing term generation.

\subsection{Syntax of \pun}
\label{sec:syntax}

\begin{figure}[t]
\begin{align*}
t ::=&\ \overline{n}\ |\ \textbf{true}\ |\ \textbf{false}\ |\ x\ |\ ()
       |\ \IF{t_0}{t_1}{t_2}\ |\ t_0 + t_1\  |\ (t_1, t_2) \\
&|\ \FST{t_0}\ |\ \SND{t_0}\ |\ \LAMBDA{x}{t_0}\ |\ t_1 t_2\
|\ \LET{x}{t_1}{t_2}\ |\ \REC x.t_0 \\
&|\ \textbf{leaf}\ |\ \NODE{t_0}{t_1}{t_2}{t_3}\ |\
    \CASE{t_0}{t_1}{p}{t_2} \\[1ex]
c ::=&\ \overline{n}\ |\ \textbf{true}\ |\ \textbf{false}\ |\ ()\ |\ (c_1, c_2)\ |\
         \LAMBDA{x}{t_0}\ |\ \textbf{leaf}\ |\ \NODE{c_0}{c_1}{c_2}{c_3}\\[1ex]
p ::=&\ c\ |\ x\ |\ \NODE{c_0}{c_1}{c_2}{c_3}\ |\ (c_1, c_2)
\end{align*}
\vspace{-6mm}
\caption{The syntax of \pun terms $t$, canonical \pun terms $c$, and \pun patterns $p$.}
\label{fig:syntax}
\label{fig:canonical}
\label{fig:patterns}
\end{figure}

The syntax of \pun is given in Figure~\ref{fig:syntax}. For terms $\overline{n}$
denotes an integer literal and \emph{x} ranges over an infinite set of
immutable \emph{variables}. You can work with terms using addition, tuples, and
the product decomposition \textbf{fst} and \textbf{snd}. Furthermore, it
contains the standard conditional, functions definition (lambda abstractions),
function application, let-expressions and recursion.

A lambda abstraction is an anonymous function defined by its input
$x$ and output $t_0$. Function application is written without parentheses
surrounding the argument, unless they are needed for syntactic disambiguation,
and is evaluated successfully on validly typed input.
The let-expression corresponds to binding the term $t_1$ to the name $x$, and
evaluating the term $t_2$ with $t1$ instead of $x$. The recursion rule
$\REC x . t_0$ substitutes into function body $t_0$ the whole term.
I.e., evaluating $\REC x . t_0$ results in $t_0 [\REC x . t_0 / x]$. Thus, this
term may be non-terminating.

Every canonical form $c$ is a well-formed
\emph{closed} term: i.e., it has no free variables and can eventually be
evaluated (or normalised) to a unique normal form.
Note in particular that the body $t_0$ of a
$\lambda$-expression only can contain $x$ as a free variable. This ensures that the
application of the abstraction to a term, eventually will result in a closed term.

Additionally, for the binary search trees,
the language has been extended
with the corresponding terms and with case-statements.
Binary search trees are implemented in the expected way:
A binary search tree is either a simple \textbf{leaf} or a \textbf{node} which
consists of a left subtree ($t_0$), a key ($t_1$), a value ($t_2$), and a right
subtree ($t_3$).

Since it has been added for the sake of a benchmark test, the implementation of case-statements is specific to binary search trees.
Its semantics corresponds to pattern-matching the term $t_0$ to either a simple
\textbf{leaf} or to a pattern $p$. Then, the appropriate term, either $t_1$ or
$t_2$, is evaluated. A pattern $p$ is a subset of the canonical terms.

\subsection{Typing of \pun}

The grammar of \pun types and typing rules for \pun are given in Figure~\ref{fig:typing-rules-pun}.
A type is either simple or compound. The simple types are integer and Boolean.
Compound types are product types (tuples), lambda abstractions (functions), and binary
search trees. In a binary search tree, two types $\tau_1$ and $\tau_2$ are the key and
value types, respectively. Finally, there is a special type \textbf{unit}, which
is interpreted in the usual way: It has only one valid term, \haskell|()|, which can
hold no further information.

The typing rules for \pun follows the conventional formalisation of a simple typed functional language (e.g. \cite{Pierce:2002:book}), with the exception of
case-statements that only are defined for binary search trees. Branching on booleans
and integers can be performed using the conditional-term. Thus, to simplify and
avoid overlap in the typing judgements, we restrict cases to binary search
trees, as that is the only term without a branching term.
The typing rules is a needed foundation as our approach is to generate well-formed, well-typed, terminating terms.

\newcolumntype{M}{>{$$}c<{$$}} 

\begin{figure}[t]
\begin{align*}
\tau ::=\ &\INT\ |\ \BOOL\ |\ (\tau_1, \tau_2)\ |\ (\tau_1 \to \tau_2)\ |\
         \textbf{BST}\ \tau_1\ \tau_2\ |\ \textbf{unit}
\end{align*}

\begin{tabular}{MM}
  \Axiom
  {Lookup}    
           {\Gamma \vdash x : \tau} $(\Gamma (x) = \tau)$
&
  \Axiom
  {Number}    
           {\Gamma \vdash \overline{n} : \INT}
 \\[1em]
  \Axiom
  {True}   
           {\Gamma \vdash \textbf{true} : \BOOL}
&
  \Axiom
  {False}   
           {\Gamma \vdash \textbf{false} : \BOOL}
\\[1em]
  \InfTwo  {\Gamma \vdash t_0 : \INT}
           {\Gamma \vdash t_1 : \INT}
  {Plus}   
           {\Gamma \vdash t_0 + t_1 : \INT}
&
  \InfTwo  {\Gamma \vdash t_0 : \INT}
           {\Gamma \vdash t_1 : \INT}
  {Leq}    
           {\Gamma \vdash t_0 \leq t_1 : \BOOL}
\\[1em]
  \InfThree {\Gamma \vdash t_0 : \BOOL}
            {\Gamma \vdash t_1 : \tau}
            {\Gamma \vdash t_2 : \tau}
  {If}      
            {\Gamma \vdash \IF{t_0}{t_1}{t_2} : \tau}
&
  \InfTwo  {\Gamma \vdash t_1 : \tau_1}
           {\Gamma \vdash t_2 : \tau_2}
  {Pair}   
           {\Gamma \vdash (t_1, t_2) : \tau_1 \times \tau_2}
\\[1em]
  \InfOne  {\Gamma \vdash t_0 : \tau_1 \times \tau_2}
  {Fst}    
           {\Gamma \vdash \FST{t_0} : \tau_1}
&
  \InfOne  {\Gamma \vdash t_0 : \tau_1 \times \tau_2}
  {Snd}    
           {\Gamma \vdash \SND{t_0} : \tau_2}
\\[1em]
  \InfOne  {\Gamma [x \mapsto \tau_1] \vdash t_0 : \tau_2}
  {Abstraction} 
           {\Gamma \vdash \LAMBDA{x}{t_0} : \tau_1 \to \tau_2}
&
  \InfTwo  {\Gamma \vdash t_1 : \tau_1 \to \tau_2}
           {\Gamma \vdash t_2 : \tau_1}
  {Application}    
           {\Gamma \vdash t_1\ t_2 : \tau_2}
\\[1em]
  \InfTwo  {\Gamma \vdash t_1 : \tau_1}
           {\Gamma [x \mapsto \tau_1] \vdash t_1 : \tau_2}
  {Let}    
           {\Gamma \vdash \LET{x}{t_1}{t_2} : \tau_2}
&
  \InfOne  {\Gamma [x \mapsto \tau] \vdash t_0 : \tau}
  {Recursion}    
           {\Gamma \vdash \REC x. t_0 : \tau}
\\[1em]
  \Axiom
  {Unit}   
           {\Gamma \vdash () : \textbf{unit}}
&
  \Axiom
  {Leaf}   
           {\Gamma \vdash \textbf{leaf}\ : \textbf{BST}\ \tau_1\ \tau_2}
\\[1em]
\multicolumn{2}{c}{
  \InfFour {\Gamma \vdash t_0 : \textbf{BST}\ \tau_1\ \tau_2}
           {\Gamma \vdash t_1 : \tau_1}
           {\Gamma \vdash t_2 : \tau_2}
           {\Gamma \vdash t_3 : \textbf{BST}\ \tau_1\ \tau_2}
  {Node}   
           {\Gamma \vdash \NODE{t_0}{t_1}{t_2}{t_3} : \textbf{BST}\ \tau_1\ \tau_2}
} \\[1em]
\multicolumn{2}{c}{
  \InfFour  {\Gamma \vdash t_0 : \textbf{BST}\ \tau_1\ \tau_2}
            {\Gamma \vdash t_1 : \tau_3}
            {\Gamma \vdash t_2 : \textbf{BST}\ \tau_1\ \tau_2}
            {\Gamma \vdash t_3 : \tau_3}
  {Case}    
            {\Gamma \vdash \CASE{t_0}{t_1}{t_2}{t_3} : \tau_3}
}
\end{tabular}
\caption{The grammar of \pun types $\tau$, and typing rules for \pun.}
\label{fig:typing-rules-pun}
\end{figure}

\section{The Generator Generator}
\label{sec:gengen}
The goal of this work is to be able to check properties without having to provide a generator used types,
and rather have them generated from \pun programs. A property consists of a
name, arguments and a term which is the property that must hold. This property is a term that must have the type boolean.
We want to substitute the arguments in the property term with generated terms of the
appropriate types. The first step is writing a generator that can generate
generators, which in turn generate terms that will be the arguments for the tests.

\subsection{Generating Generators}
In the addition example, we need a generator
for integer terms, because the addition operator in \pun only is defined for
integers. In the case of the binary search tree functions, we can look to the
key and value types to know what types are valid input types.
For a well-defined function, it is in general possible to infer the types of the
required generators directly from the program. The more interesting task, then,
is to write a function \haskell|generateGenerator|.

The function should be able to generate a generator for
arbitrary well-formed and well-typed terms of the requested type that can be
evaluated when substituted into a property. To do so, it
needs to know the requested type, but also information about existing name
bindings and types in the program. \haskell|generateGenerator| needs to know the
global name bindings so it can generate arbitrary names for function
abstractions, let-statements and recursive terms without binding a name that already
exists on the top-level of the program. These would cause conflict upon
evaluation.

It also needs to know what names have been bound to which types in
the scope of the function, for two reasons. First, so that it may avoid
conflicts when using names inside function abstractions, let-statements and recursive
terms, since these names are not relevant outside the scope of each term. I.e.,
in the term $(\lambda x . x + 5)$, the name $x$ can be used freely
outside the function abstraction and other bindings to that name are not
relevant within the scope of the function. Second, so that it may with some
probability generate a term of a type that is already inhabited and bound to a
name elsewhere in the program.

When writing \haskell|generateGenerator|, a significant challenge was how to
generate \emph{complex} terms of a given type. I.e.,
we wish to conceive a strategy for generating terms of various complexities and
types that can be \emph{evaluated} to a term of the correct type. For example, a
valid term of type integer might be $(\lambda x . x + 5)\ 3$, since it normalises to
$8$. However, in attempting to generate more interesting terms, we may also generate
some terms that are ill-typed, ill-formed, or non-terminating.

Our approach to this problem is influenced by Fetscher et
al.~\cite{Fetscher:2015} in their paper on generating well-typed terms from type
systems. The requested type to be generated, dictates which typing rules could
have resulted in such a term. We therefore start with a goal term on the form $\Gamma
\vdash t_0 : \tau_0$, meaning that from the program environment environment $\Gamma$, we wish to generate a
term $t_0$ of the requested type $\tau_0$. We then work our way upwards and
generate arbitrary derivations that result in this goal.

To generate the derivations, we must look to our typing rules to see which of
them could have resulted in a term of type $\tau_0$. For the sake of argument, let
us say we wish to generate a generator
for pairs of type $(\BOOL \times \INT)$. Then there are multiple
appropriate rules. However, we may also notice that some of the applicable
rules, require us to generate another, \emph{larger} term. For instance, we
could have generated another pair term $t_1 : ((\BOOL \times
\INT) \times\INT)$, and then applied the \textbf{fst}-rule to get a valid result $\textbf{fst}(t_1) : ((\BOOL \times
\INT) \times\INT)$. But this way, we might end up generating bigger
and bigger terms indefinitely, e.g., a term of type $(\tau_0 \times (\tau_1 \times (\tau_2 \times ...)))$. Therefore, we have to be careful about how we
select rules. Some rules, such as the axiomatic rules, result in a well-typed, but smaller term.

To bound the recursion of \haskell|generateGenerator|, we use the
\quickcheck combinator \haskell|sized|. It constructs a generator that depends
on the \haskell|size| parameter, which is passed
from \quickcheck to the generator. Initially, \quickcheck generates small test
cases and increases the size to generate more complex test cases as testing progresses. In our case,
the parameter controls the depth of the recursion and therefore of the
derivations. When we have also disqualified some rules to begin with, this is a solution to bounding recursion of the term
\emph{generators}. We discuss the issue of recursive \emph{terms} in Section~\ref{sec:interesting-terms}.

By selecting amongst the applicable rules randomly, while being careful to
ensure termination, we can generate random derivations and guarantee the result
is a well-typed, well-formed term. The strategy used by
\haskell|generateGenerator| in our implementation of \pun is summarised
in Figure~\ref{fig:gen-rules}. The square brackets represent the generator
function, and its argument is the requested type. The number $n$ represents the size of the term we are
currently generating. The left-hand side of the rule represents the rule we are
applying in the next step of the generation.
$\text{arbitrary}$ is an auxiliary function that selects an arbitrary element
from the given set. Note that generating a new variable $x$ means either finding an
existing variable in the environment with that name or generating a new term of
that type. An unbound variable is uninteresting to generate.

\begin{figure}[t]
\text{For arbitrary types $\tau$, $m < n$, }
\begin{align*}
\GEN{\INT}{0}\ &=\ arbitrary(\mathbb{N}) \\
\GEN{\INT}{n}\ &=\ \begin{cases}
      \GEN{\INT}{m} + \GEN{\INT}{m}, &\\
      \IF{\GEN{\BOOL}{m}}{\GEN{\INT}{m}}{\GEN{\INT}{m}}, & \\
      \FST{\GEN{(\INT, \tau)}{m}}, & \\
      \SND{\GEN{(\tau, \INT)}{m}}, & \\
      (\LAMBDA{\GEN{\tau}{m}}{\GEN{\INT}{m}}) \GEN{\tau}{m}, & \\
      \LET{\GEN{\tau}{m}}{\GEN{\tau}{m}}{\GEN{\INT}{m}}, & \\
      \REC \GEN{\INT}{m} . \GEN{\INT}{m}, & \\
      \GEN{(x : \INT)}{m} &
\end{cases} \\
\GEN{\BOOL}{0}\ &=\ arbitrary(\{\textbf{true}, \textbf{false}\}) \\
\GEN{\BOOL}{n}\ &=\ \begin{cases}
      \GEN{\BOOL}{m} \leq \GEN{\BOOL}{m}, &\\
      \IF{\GEN{\BOOL}{m}}{\GEN{\BOOL}{m}}{\GEN{\BOOL}{m}}, & \\
      \FST{\GEN{(\BOOL, \tau)}{m}}, & \\
      \SND{\GEN{(\tau, \BOOL)}{m}}, & \\
      (\LAMBDA{\GEN{\tau}{m}}{\GEN{\BOOL}{m}}) \GEN{\tau}{m}, & \\
      \LET{\GEN{\tau}{m}}{\GEN{\tau}{m}}{\GEN{\BOOL}{m}}, & \\
      \REC \GEN{\BOOL}{m} . \GEN{\BOOL}{m}, & \\
      \GEN{(x : \BOOL)}{m} &
\end{cases} \\
\GEN{(\tau_1, \tau_2)}{n}\ &=\ (\GEN{\tau_1}{m}, \GEN{\tau_2}{m}) \\
\GEN{(\tau_1 \to \tau_2)}{n}\ &=\ \GEN{\tau_1}{m} \to \GEN{\tau_1}{m} \\
\GEN{\textbf{BST}\ \tau_1\ \tau_2}{n}\ &=\ \begin{cases}
      arbitrary(\{\textbf{leaf}\}) & \\
      \NODE{\GEN{\textbf{BST}\ \tau_1\ \tau_2}{m}}{\GEN{\tau_1}{m}}{\GEN{\tau_2}{m}}{\GEN{\textbf{BST}\ \tau_1\ \tau_2}{m}}
\end{cases}
\end{align*}
\vspace{-6mm}
\caption{Generator rules for \pun.}
\label{fig:gen-rules}
\end{figure}

\section{Generating Interesting Terms is Hard}
\label{sec:interesting-terms}

With a type checker, an interpreter and a term generator generator in hand,
writing a function that checks a property several times is, surprisingly
straight forward. Essentially, the \haskell|property| statement consists of
a list of variables and a term to evaluate. Substituting in terms of the
correct type is a matter of inferring a type for each variable, and then
generating a term of that type. The rest is finicky book keeping, that keeps
track of the number of tests that have been performed etc (for error
messages).

However, while writing type checkers and interpreters is a well explored
area of research, writing generator generators remains a novel an
challenging endeavour. On one hand, when generating primitively typed terms
we run the risk of generating terms that are too "boring". For instance, we
may generate a simple numeric \pun terms for the type \INT, even though \INT
can also be the result of a term was typeable by the Application rule from
Figure~\ref{fig:typing-rules-pun}. Helping the programmers that wants to
test their program thoroughly, can be to come up with terms of different
complexity, because those are more interesting to test the program with.  We
have made some decisions for recursive, lambda and let terms with regards to
what the user would do if written by hand.

On the other hand, we also face the challenge of bounding the generated
terms. For instance, for terms of type \INT, we could generate terms such as
\begin{align*}
\REC{}x.~x + 1
\end{align*}
\noindent
that do not terminate.

The current implementation of \pun limits such recursion in a very
conservative way to ensure that terms are terminating. For instance, for
\texttt{rec} we do not allow the bound variable to appear in its body at
all, and since there is no other recursion possible in the language, never
generating recursive terms ensures that all generated terms terminate.  This
strictness can be relaxed to recursion schemes that are known to terminate,
but it has been left as future work. Together with an approach that generate
terms that terminate by the size-change termination principle~\cite{Jones:2001}.

Where defining what ``smaller'' means with regards to terms is
non-trivial in the general case. Instead, we have made the conservative choice
to only allow the use of the \REC rule when the argument to the rule $f$ is a
function $f : a \to b$ such that the function argument $a$ approaches a
terminating base case for each call. Thus, the program below
is a valid \pun term containing a recursive function definition and call, since
the argument to the \REC rule is a function, whose argument $n$
gets smaller, and thus closer to the base case, for each call.

\begin{lstlisting}[escapeinside=\%\%]
let fib = rec f . (\n . if   n <= 1
                        then 1
                        else f (n - 1) + f (n - 2))
in  fib 5
\end{lstlisting}

\noindent
The above restriction certainly disallows some interesting recursive terms, but
it still allows for a class of interesting recursive functions that we can
be sure will terminate.

\subsection{Function Abstractions and Let-Statements}
When generating a function abstraction or let-statement, we have taken steps to
increase probability that the program will use the generated variable name in
the body of the term.
This is because it
seems pointless for the program to generate names that are not used. If
a programmer were to write a test program, they would likely define one of those
terms with the intention of using the names.
For a programming language with side-effects it would make sense to have a let-statement
that is only used for computing something with side-effects. This could for instance
be an assertion or a print for debugging. Once the programmer is done debugging, these
terms can easily be commented out since they are now not used.
In \pun there are no side effects, so there is no reason to have a let-statement or
abstraction, where the variable name is not used in the term body.

The types of the terms generated inside the body of a let-statement or function abstraction, are randomly
generated. Therefore, it is not sufficient to only increase the probability of choosing
a variable. If the let-term is of type integer and the body never has to generate
something of type integer, then the name will not be used. We therefore altered the program
further for the type generator to have a higher probability of generating a type
that can be found in the bindings, which then allows for a variable to be generated.

\section{Evaluation}
\label{sec:evaluation}

We will now evaluate \pun by benchmarking it against a Haskell
implementation of binary search trees as Hughes describes in~\cite{Hughes:2019}.
The Haskell implementation is used as a benchmark for \pun. The properties
tested when providing a generator, like Hughes does for QuickCheck, should
yield the same result when written in \pun. This means that when we introduce
an error in the Haskell implementation and the equivalent error in \pun, then
\pun should provide an error message as well. The bencmark suite therefore includes
several faulty properties, as Hughes describes in the paper, to check that \pun
does in fact find the same errors as \quickcheck does for the Haskell implementation.
Using Hughes' examples exposes \pun for several kinds of properties;
validity, metamorphic, inductive, model-based, post-condition and
preservation of equivalence. This way we can check that \pun finds the same
errors as QuickCheck regardless of the kind of property that is being checked.

Part of the result is checking whether \pun finds a bug that has been planted,
another part is measuring how much work is saved by not having to provide a
generator. The amount of work saved is measured in lines of code.
We discuss comparative results at the end of the paper.

For metamorphic properties, Hughes checks whether two trees have the same
content by first converting them to lists and then comparing them. Because
\pun does not have built-in lists, we use a particular binary search tree
as a way of representing lists when defining our properties in \pun.
We have a function written in \pun that takes a binary search tree and modifies
it to a binary search tree that is isomorphic to lists. The regular trees
contain a key and a value as separate information about the tree, whereas the
modified tree contains both the key and value as a tuple where only the key
would originally have been stored. The value is replaced by unit, which is
used as a way of representing that there is no extra information there.
All the left sub trees are replaced with a \emph{leaf} which does not add
any new information. The tree is built from left to right, so any new node appears
as a right sub tree.
Below to is an example of a tree with the regular binary search tree
structure.

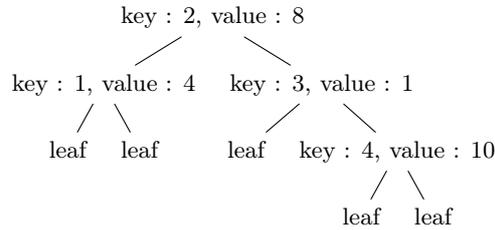
\begin{center}
\small
\begin{forest}
  for tree={
      grow=south
  }
  [{key : 2, value : 8}
    [{key : 1, value : 4}
      [leaf]
      [leaf]
    ]
    [{key : 3, value : 1}
      [leaf]
      [{key : 4, value : 10}
        [leaf]
        [leaf]
      ]
    ]
  ]
\end{forest}
\end{center}

\noindent
After using the function \haskell|model|, we end up with the following same tree
but with the structure of the model that we have chosen for lists.

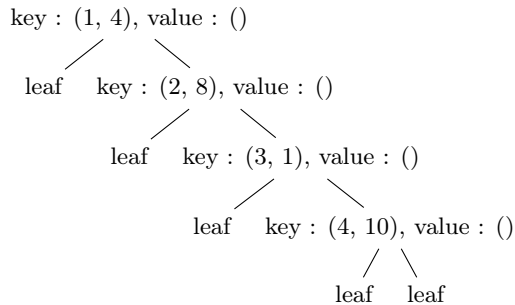
\begin{center}
\small
\begin{forest}
  for tree={
    grow=south,
  }
  [{key : (1, 4), value : ()}
    [leaf]
    [{key : (2, 8), value : ()}
      [leaf]
      [{key : (3, 1), value : ()}
        [leaf]
        [{key : (4, 10), value : ()}
          [leaf]
          [leaf]
        ]
      ]
    ]
  ]
\end{forest}
\end{center}

Hughes also uses lists as a model for the model-based properties, in which
case we also use the same function that turns a regular binary search tree
to one that is isomorphic with lists.
The reason this structure is isomorphic is because there are equally many elements
in the tree as there would have been in a list. Each key that contains a tuple
\emph{(key, value)} counts as an element. There are also mappings in both directions,
such that you can have a list and represent it as this specific binary tree
structure, and also be able to go from that tree back to its original list form.

Initial bench marking shows that at least blatant errors are easily detected
by \pun, though it may run more tests before finding a counter example.  For
instance, when changing the implementation of \haskell|insert| from
Section~\ref{sec:approach} to
\begin{lstlisting}[escapeinside=\%\%]
insert k1 v1 t = [node leaf k1 v1 leaf].
\end{lstlisting}
\noindent and
\begin{lstlisting}[escapeinside=\%\%]
insert k v _ = Branch Leaf k v Leaf
\end{lstlisting}
\noindent respectively.
The property \haskell|find-post-present| usually fails within 50 tests in
both implementations.

\section{Conclusion}
\label{sec:conclusion}

In this paper we have proposed the programming language \pun that is a
higher order functional language extended with properties. Based on the
formalisation of the language, we have shown how we from well-typed programs
can generate \quickcheck generators. We have added limitations to the
generation of terms, to ensure that the test will terminate.  To show that
\pun can find the same bugs in a program as a Haskell implementation with
QuickCheck, we extended the language with binary search trees.  The
evaluation showed that we could find the same errors in many cases, but that
some one sometimes need a transformation function (such as
\haskell|validify| in Section~\ref{sec:approach}), or that the tests may
need to run more times.

In the future we would like to extend with more data types like lists, which
would make it easier to define certain properties, and even generalised to
general algebraic data types, that the programmer can define themselves.  It
will make \pun programs that are easier implement and improve readability,
which in turn makes the code easier to maintain.  We would also need to
iterate on the heuristics for interesting terms to generate a larger class
of terms that will result in terminating tests.
And finally, investigate if reverse interpretation of the property
generation can improve the
accuracy~\cite{KristensenEtal:2022:NIK,ThomsenAxelsen:2016:IFL}.

\bibliographystyle{abbrv}
\bibliography{mybibliography}

\begin{thebibliography}{10}

\bibitem{ClaessenHughes:2000:ACM}
K.~Claessen and J.~Hughes.
\newblock Quickcheck: A lightweight tool for random testing of {Haskell}
  programs.
\newblock In {\em Proceedings of the Fifth ACM SIGPLAN International Conference
  on Functional Programming}, ICFP '00, page 268–279. ACM, 2000.

\bibitem{Fetscher:2015}
B.~Fetscher, K.~Claessen, M.~Pa{\l}ka, J.~Hughes, and R.~B. Findler.
\newblock Making random judgments: Automatically generating well-typed terms
  from the definition of a type-system.
\newblock In {\em Programming Languages and Systems}, pages 383--405. Springer
  Berlin Heidelberg, 2015.

\bibitem{Graham:2022}
D.~Graham, R.~Black, and E.~v. Veenendal.
\newblock {\em Foundations of Software Testing: ISTQB Certification}.
\newblock Cengage Learning EMEA, 4 edition, 2022.

\bibitem{Hughes:2016}
J.~Hughes.
\newblock Experiences with {QuickCheck}: Testing the hard stuff and staying
  sane.
\newblock In {\em A List of Successes That Can Change the World}, pages
  169--186. Springer International Publishing, 2016.

\bibitem{Hughes:2019}
J.~Hughes.
\newblock How to specify it! a guide to writing properties of pure functions.
\newblock In {\em Trends in Functional Programming: 20th International
  Symposium, TFP 2019}, page 58–83. Springer-Verlag, 2019.

\bibitem{KristensenEtal:2022:NIK}
J.~T. Kristensen, R.~Kaarsgaard, and M.~K. Thomsen.
\newblock Branching execution symmetry in {Jeopardy} by available implicit
  arguments analysis.
\newblock In A.~Rutle, editor, {\em Proceedings of 34th Norwegian ICT
  Conference for Research and Education, NIKT 2022}, number~1, 2022.

\bibitem{LampropoulosEtal:2017}
L.~Lampropoulos, D.~Gallois-Wong, C.~Hri\c{t}cu, J.~Hughes, B.~C. Pierce, and
  L.-y. Xia.
\newblock Beginner's luck: A language for property-based generators.
\newblock In {\em Proceedings of the 44th ACM SIGPLAN Symposium on Principles
  of Programming Languages}, POPL '17, page 114–129. ACM, 2017.

\bibitem{Jones:2001}
C.~S. Lee, N.~D. Jones, and A.~M. Ben-Amram.
\newblock The size-change principle for program termination.
\newblock {\em SIGPLAN Not.}, 36(3):81–92, 2001.

\bibitem{Pierce:2002:book}
B.~C. Pierce.
\newblock {\em Types and Programming Languages}.
\newblock The MIT Press, 2002.

\bibitem{ThomsenAxelsen:2016:IFL}
M.~K. Thomsen and H.~B. Axelsen.
\newblock Interpretation and programming of the reversible functional language.
\newblock In {\em Proceedings of the 27th Symposium on the Implementation and
  Application of Functional Programming Languages}, IFL '15, pages 8:1--8:13.
  ACM, 2016.

\bibitem{Whittaker:2000}
J.~Whittaker.
\newblock What is software testing? {A}nd why is it so hard?
\newblock {\em IEEE Software}, 17(1):70--79, 2000.

\bibitem{Winskel:1993}
G.~Winskel.
\newblock {\em The Formal Semantics of Programming Languages : an
  introduction}.
\newblock The MIT Press, 1993.

\end{thebibliography}

\end{document}